\documentclass[12pt,preprint]{aastex}

\shorttitle{The Sun's Shallow Meridional Circulation}
\shortauthors{Hathaway}

\begin{document}


\title{The Sun's Shallow Meridional Circulation}
\author{David H. Hathaway}
\affil{NASA Marshall Space Flight Center, Huntsville, AL 35812 USA}
\email{david.hathaway@nasa.gov}

\begin{abstract}
The Sun's global meridional circulation is evident as a slow poleward flow at its surface. This flow is observed to carry magnetic elements poleward - producing the Sun's polar magnetic fields as a key part of the 11-year sunspot cycle. Current theories for the sunspot cycle assume that this surface flow is part of a circulation which sinks inward at the poles and turns equatorward at depths below 100 Mm. Here we use the advection of the Sun's convection cells by the meridional flow to map the flow velocity in latitude and depth. Our measurements show the largest cells clearly moving equatorward at depths below 35 Mm - the base of the Sun's surface shear layer. This surprisingly shallow return flow indicates the need for substantial revisions to solar/stellar dynamo theory.
\end{abstract}


\section{INTRODUCTION}

The Sun's meridional circulation has been a part of solar magnetic dynamo theory for half a century. A poleward meridional flow from the Sun's mid latitudes was invoked in the earliest models (long before the flow was actually measured) to transport magnetic elements from decaying sunspot regions to the poles where they would erode the opposite polarity magnetic field from the old sunspot cycle and build up the polar fields of the new sunspot cycle \citep{Babcock61}. This surface meridional flow, along with its latitudinal structure and variation in time, is now well observed \citep{Topka_etal82, Komm_etal93, Hathaway_etal96, HathawayRightmire10, HathawayRightmire11} and its role in the surface magnetic flux transport is well established \citep{DeVoreSheeley87, vanBallegooijen_etal98, SchrijverTitle01}. Dynamo theories over the last decade and a half have assumed that the mass traveling poleward in the surface layers sinks inward at the poles and returns to the equator along the base of the Sun's convection zone at a depth of 200 Mm. In these theories this slow, dense, equatorward flow is responsible for the equatorward drift of sunspot activity \citep{DikpatiChoudhuri94, DikpatiCharbonneau99, NandyChoudhuri02}. Dynamo models based on this deep meridional circulation have recently been used to predict the currently emerging sunspot cycle, albeit with disparate results \citep{Dikpati_etal06, Choudhuri_etal07}.

The meridional flow is difficult to measure. Its amplitude ($\sim 10-20 \rm{\ m\ s}^{-1}$) is more than an order of magnitude weaker than that of the other major flows observed on the surface of the Sun. The axisymmetric longitudinal flow, differential rotation, has a dynamic range of $\sim 200 \rm{\ m\ s}^{-1}$ and the non-axisymmetric cellular convection flows have typical velocities of several hundred $\rm{\ m\ s}^{-1}$. The meridional flow is observed to be poleward from the equator with peak flow speeds in the mid latitudes. The flow amplitude measured from Doppler shifts of spectral lines formed at the surface is $\sim 20 \rm{\ m\ s}^{-1}$ \citep{Hathaway_etal96, Ulrich10} while the amplitude found by measuring the motions of the small magnetic features is $\sim 12 \rm{\ m\ s}^{-1}$ \citep{Komm_etal93, HathawayRightmire10, HathawayRightmire11}. The meridional flow can also be measured using the methods of local helioseismology which yield a peak velocity of $\sim 20 \rm{\ m\ s}^{-1}$ that appears to be constant with depth down to $\sim 26$ Mm \citep{Giles_etal97, SchouBogart98}. Recently, however, two new measurement methods have indicated a decrease in amplitude with depth. A method using global helioseismology \citep{Mitra-KraevThompson07} found a meridional flow that decreased with depth and became equatorward at a depth of only 40 Mm - but with a large range of error. A method using the movement of the larger solar convection cells, supergranules, also found a meridional flow that decreased with depth but without precise depth information and without detection of a return flow \citep{Hathaway_etal10}. While some helioseismic studies indicate a poleward meridional flow at depths well below 26 Mm, \cite{DuvallHanasoge09} found that those methods are prone to systematic errors and \cite{GoughHindman10} conclude that the flow below 30 Mm remains unknown. Furthermore \cite{Beckers07} has suggested that projection effects may have also compromised some of the local helioseismology results and concludes that the meridional flow velocity may decrease with depth.

Here we measure the meridional flow by tracking the motions of supergranules, but extend the analysis to include larger cells with deeper roots. Numerical models of compressible convection with radiative transfer in the near surface layers \citep{SteinNordlund00} clearly show that small cells dominate at the surface and larger structures are found at increasing depth. Supergranules cover the surface of the Sun and have a broad range of sizes that sample a corresponding range of depths. We measure the motion of the pattern of supergranules by analyzing data acquired by the Michelson Doppler Imager (MDI) \citep{Scherrer_etal95} on the ESA/NASA Solar and Heliospheric Observatory (SOHO) satellite in 1996 and 1997.

\section{DATA PREPARATION}

The data consist of 1024x1024 pixel images of the line-of-sight velocity determined from the Doppler shift of a spectral line due to the trace element nickel in the solar atmosphere. The images are acquired at a 1 min cadence. We average them over 31 min with a Gaussian weighting function which filters out any velocity components that vary on time scales less than about 16 min. We then map these temporally filtered images onto a 1024x1024 grid in heliographic latitude from pole to pole and in longitude $\pm90\degr$ from the central meridian (Figure 1). This mapping accounts for the position angle of the Sun's rotation axis relative to the imaging CCD and the tilt angle of the Sun's rotation axis toward or away from the spacecraft. Both of these angles include modification for the most recent determinations of the orientation of the Sun's rotation axis \citep{BeckGiles05, HathawayRightmire10}. We analyze data obtained during two 60+ day periods of continuous coverage - one in 1996 from May 24 to July 24 and another in 1997 from April 14 to June 18.

We also generate and analyze simulated data to assist in our determination of the representative depths. We construct the simulated data from an evolving spectrum of vector spherical harmonics in such a manner as to reproduce the spatial, spectral, and temporal behavior of the observed cellular flows \citep{Hathaway_etal10}. The cells are advected in longitude by differential rotation and in latitude by meridional flow, both of which vary with cell size.

\begin{figure}[ht]
\centerline{\includegraphics[width=0.8\textwidth]{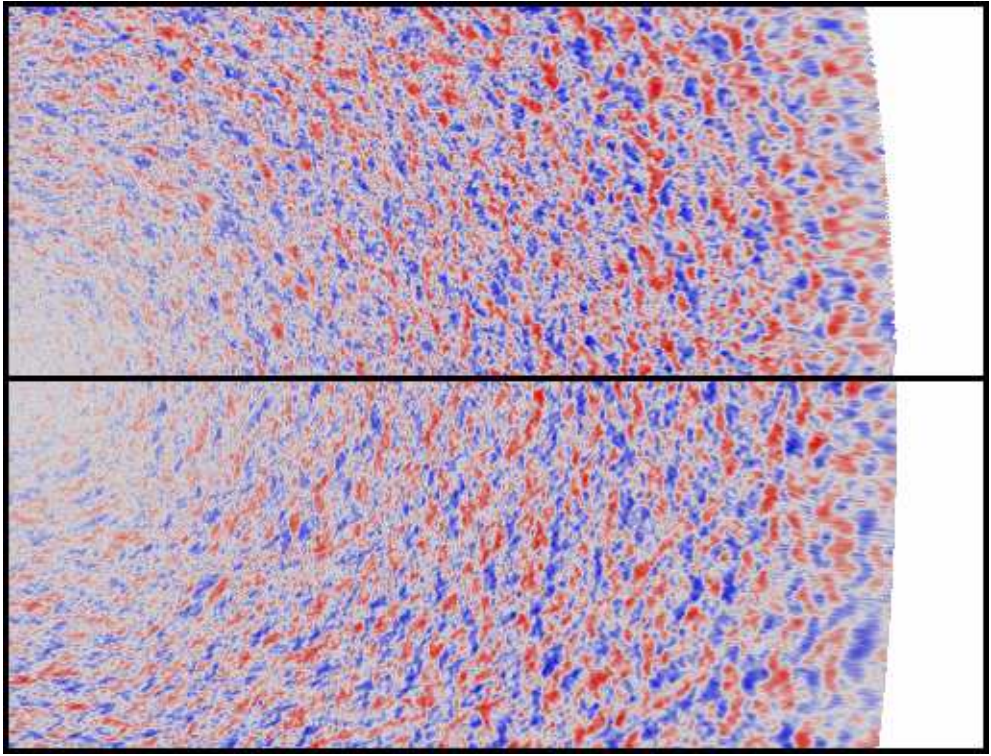}}
\caption{
Heliographic map details of the line-of-sight (Doppler) velocity from SOHO/MDI (top) and from the data simulation (bottom). Each map detail extends $90\degr$ in longitude from the central meridian on the left and about $35\degr$ in latitude from the equator (the thick horizontal line). The mottled pattern is the Doppler signal (blue for blue shifts and red for red shifts) due to the supergranule convection cells. The latitudinal movement of these supergranules yields a measurement of the Sun's meridional circulation.}
\end{figure}

\section{DATA ANALYSIS AND RESULTS}

We determine the motions of the cellular patterns in longitude and latitude for strips of data by finding the displacement of the maximum in the cross-correlation with similar strips from images acquired at time lags of 2, 4, 8, 16, 24, and 32 hr. Each strip is 11 pixels or $\sim 2\degr$ high in latitude and 600 pixels or $\sim 105\degr$ long in longitude. We repeat this procedure for 796 latitude positions between $\pm 70\degr$ latitude and for each hour over the 60 days of each MDI dataset and 60 days of simulated data. This cross-correlation method was first used to determine the equatorial rotation rate by \cite{Duvall80} who concluded that larger cells live longer and rotated faster. \cite{BeckSchou00} used a 2D Fourier transform method and found a rotation rate that increased with the wavelength of the features.

We calculate the average differential rotation and meridional flow profiles and fit them with 4th order polynomials in $\sin \theta$ where $\theta$ is the heliographic latitude (Figure 2). The rotational velocity increases with increasing time lags to a maximum at 24 hr but then decreases at 32 hr. The meridional flow velocity decreases with time lag and, at the 32 hr time lag, reverses sign. The cellular flows that live long enough to be positively correlated 32 hrs later are moving equatorward. This is a clear detection of the meridional return flow. The individual data points have a standard error of $\sim 1 \rm{\ m\ s}^{-1}$ but the vast majority of points indicate an equatorward flow significantly bigger than this. The curves fit through the data points indicate an equatorward return flow of $1.8 \rm{\ m\ s}^{-1}$ with a standard error of $< 0.2 \rm{\ m\ s}^{-1}$.

\begin{figure}[ht]
\centerline{\includegraphics[width=0.9\textwidth]{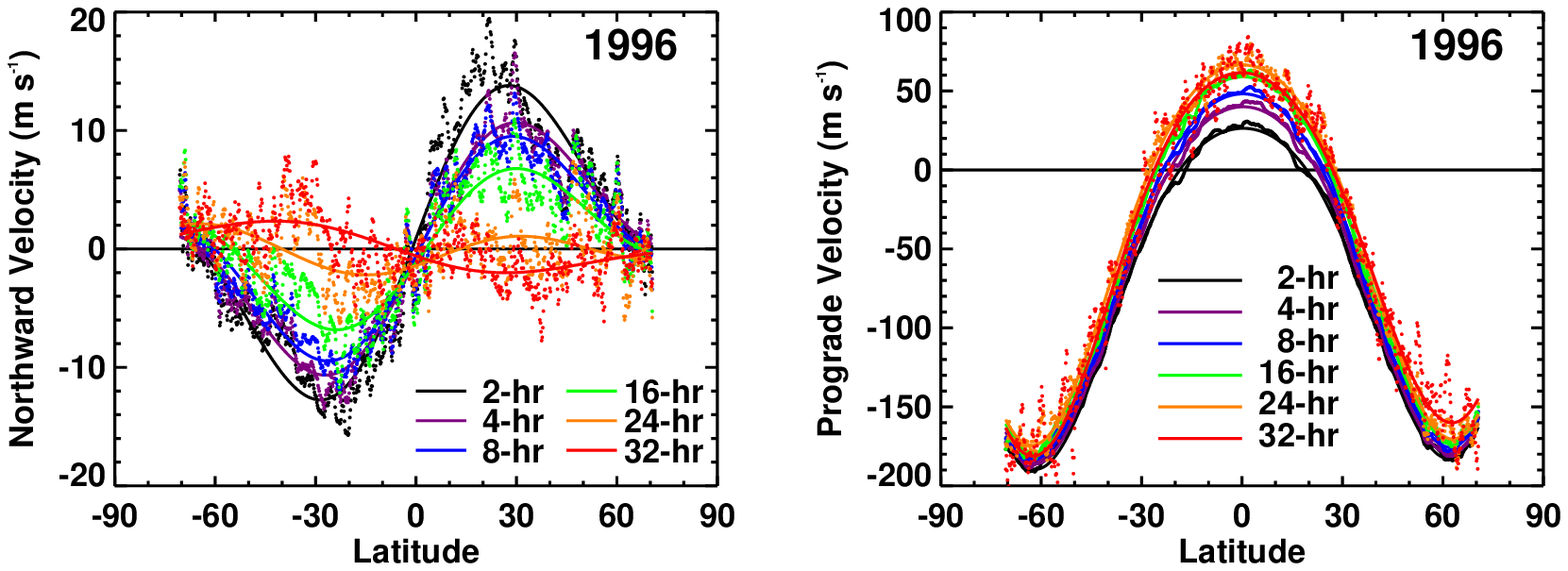}}
\centerline{\includegraphics[width=0.9\textwidth]{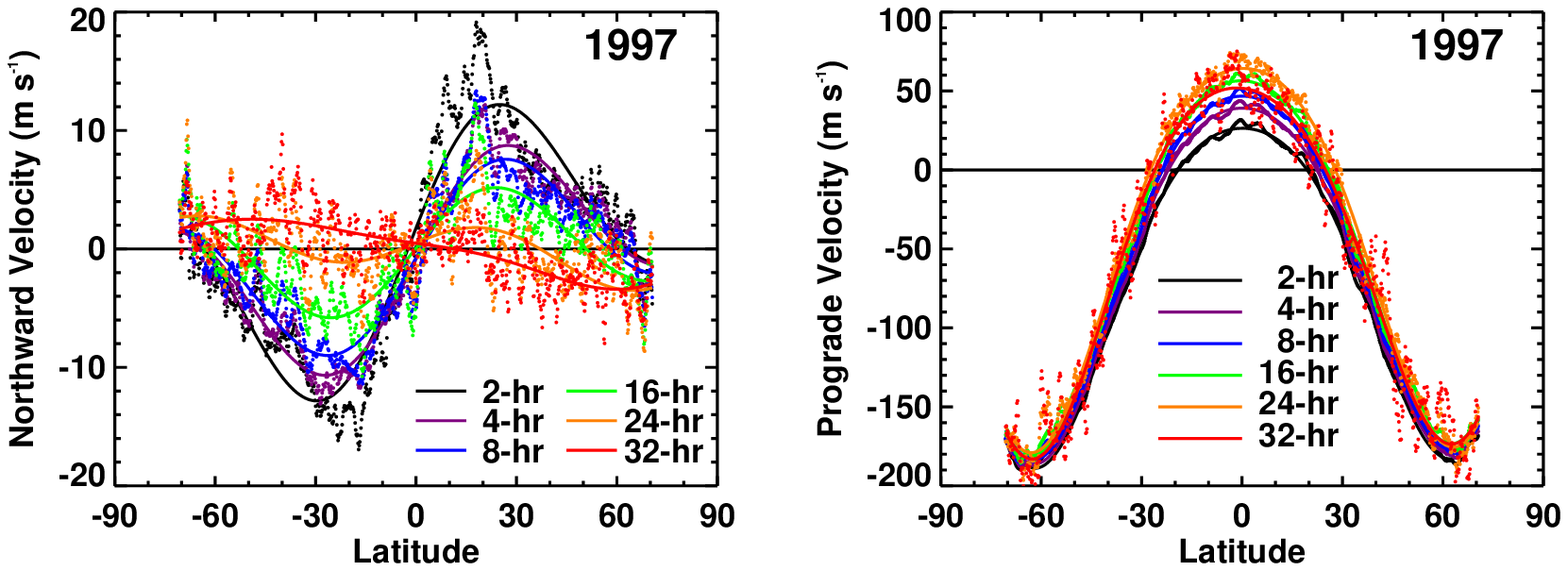}}
\centerline{\includegraphics[width=0.9\textwidth]{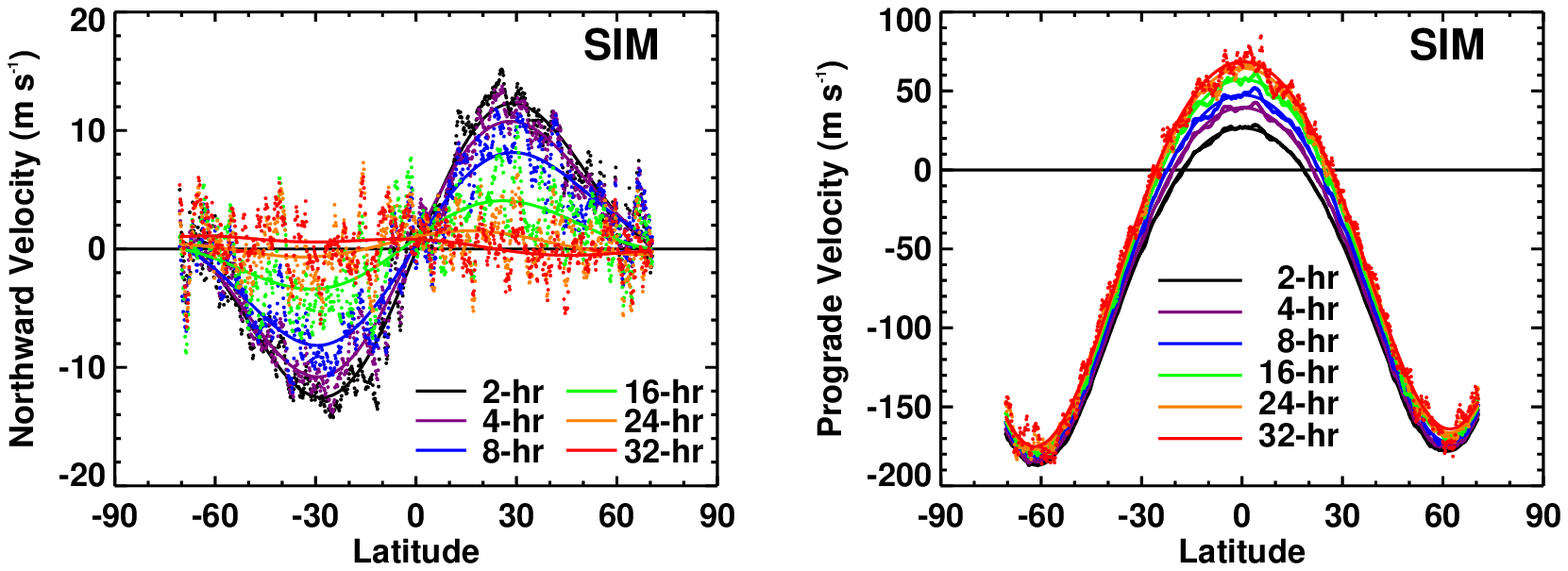}}
\caption{
Flow profiles as functions of latitude from the 1996 SOHO/MDI data (top row), the 1997 SOHO/MDI data (middle row), and the simulated data (bottom row). The flow velocities measured at each latitude are shown with colored dots for each time lag as indicated in the figure. The solid lines with the same color coding represent the 4th order polynomial fits to each profile. The meridional flow decreases in amplitude with increasing time lag and reverses direction for 32 hr lags. The rotation rate increases as the time lag increases up to 24 hr then drops at 32 hr.}
\end{figure}

We determine the characteristic convection cell wavelengths for the different time lags using the wavelength dependence of the differential rotation and meridional flow profiles used in the simulation. The differential rotation measurements are largely reproduced with a relatively simple rotation velocity, $u$, relative to an inertial (sidereal) frame of reference given by

\begin{eqnarray}
u(\theta,\lambda) & = & [(1980 - 246 \sin^2 \theta - 365 \sin^4 \theta) \cos \theta] \nonumber \\
& & [1.0 + 0.046 \tanh(\lambda/35)] \rm{\ m\ s}^{-1}
\end{eqnarray}

\noindent
where the wavelength, $\lambda$, is given in Mm. (Note that the prograde velocities plotted in Figure 2 are relative to a frame of reference rotating at the Carrington rotation rate with $u_C(\theta) = 1991 \cos \theta$.) The meridional flow measurements are largely reproduced with a northward velocity, $v$, given by

\begin{eqnarray}
v(\theta,\lambda) & = & [(65 \sin \theta - 78 \sin^3 \theta) \cos \theta] \nonumber \\
& & [\tanh((35 - \lambda)/20)] \rm{\ m\ s}^{-1}
\end{eqnarray}

\noindent
where the reproduction of the return flow is particularly sensitive to the zero crossing occurring at $\sim 35$ Mm.

Figure 3 shows the equatorial differential rotation velocity relative to the surface and the meridional flow velocity at $30\degr$ latitude along with the data points from the MDI (averaged for the two years) and simulation data analyses. The data points from the simulation virtually coincide with those from the MDI data except at the 32 hr time lag and for the northward velocity at the 16 hr time lag. No doubt, better fits could be obtained with more complicated flow profiles. It is apparent, however, that the differential rotation velocity must decrease for wavelengths greater than $\sim 35$ Mm and that the reversal in the meridional flow direction must be more abrupt.

Figure 3 also shows that the points do not fall right on the curves for the input flow profiles. We attribute this to two different processes. The two latitudes chosen for this figure represent the latitudes at which each flow reaches its maximum. Since the convection cells span a finite range of latitudes the measured values should be less than these maxima. However, the differential rotation signal is subject to an additional line-of-sight projection effect \citep{Hathaway_etal06} which makes the Doppler features appear to rotate faster. This effect raises the measured values to increasingly higher values with increasing wavelength for the prograde velocity. 

\begin{figure}[ht]
\centerline{\includegraphics[width=0.8\textwidth]{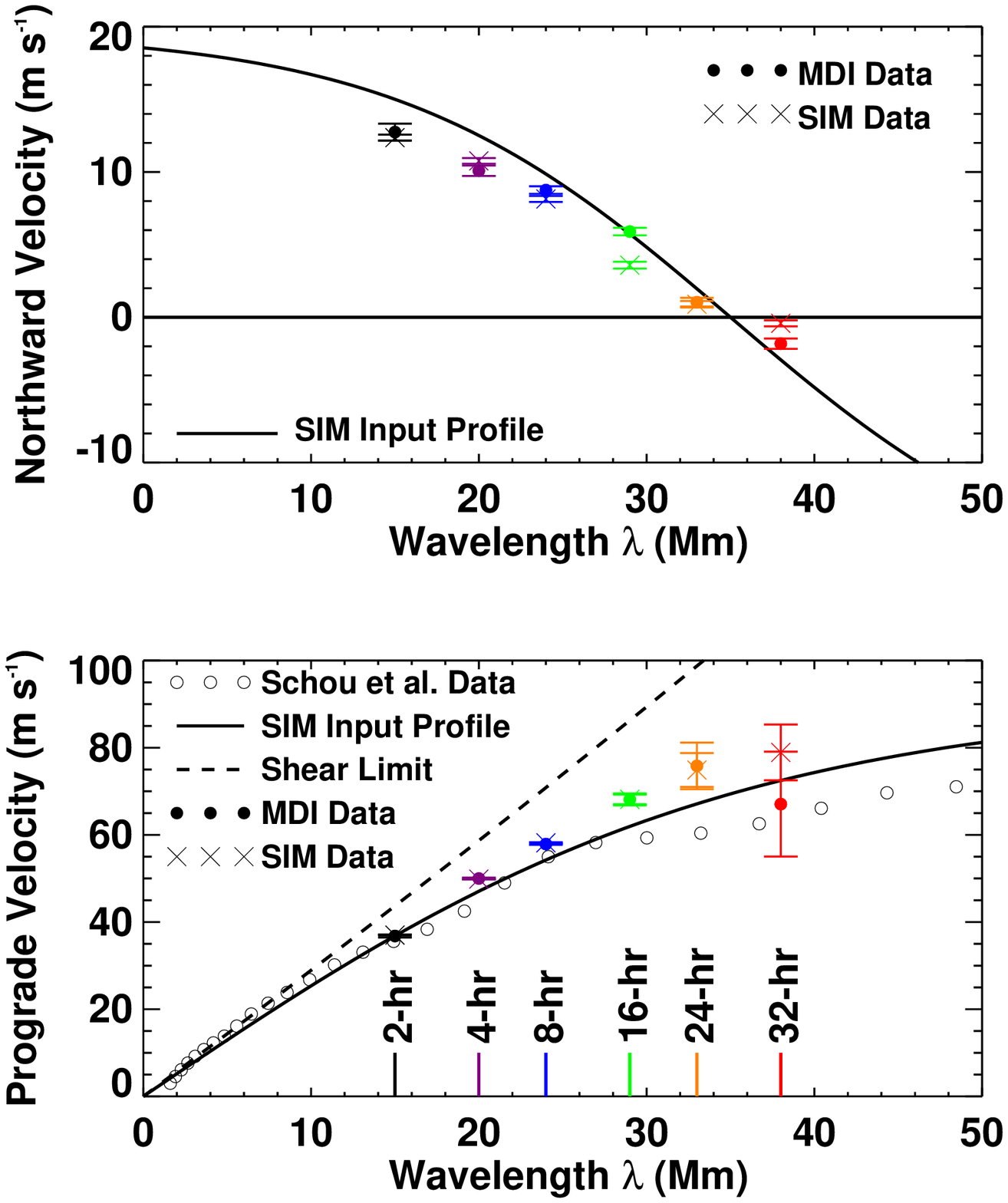}}
\caption{
The simulation meridional flow speed at $30\degr$ latitude (top) and equatorial differential rotation relative to the surface (bottom) as functions of wavelength are shown by the solid lines. The observed values from the cross-correlation analysis with the MDI data and the simulated data (filled circles and crosses respectively) are shown at their characteristic wavelengths. Error bars centered on each symbol represent $2\sigma$ errors.  The dashed line shows the theoretical limit to the rate of increase in rotation rate at the equator and the open circles show the rotation velocity determined from global helioseismology by \cite{Schou_etal98} both assuming that the wavelength equals the anchoring depth of the cells.}
\end{figure}

\section{CONCLUSIONS}

The relationship between the wavelength of a convection cell and the depth at which it is anchored or steered is well constrained by the stability of the surface shear layer and observations of the rotation rate with depth from global helioiseismology. An increase in rotation rate with depth has long been suggested by observations and is attributed to the conservation of angular momentum for fluid elements moving inward and outward in the near surface layers \citep{FoukalJokipii75, GilmanFoukal79, Hathaway82}. However, a rotation rate which increases inward faster than that given by the conservation of angular momentum is dynamically unstable \citep{Chandrasekhar61}. Measurements of this rotation rate increase from helioseismology \citep{Schou_etal98, CorbardThompson02} indicate that it follows this critical gradient to depths of 10-15 Mm. This gradient (the dashed line in Figure 3) and the helioseismicly determined rotation rates (open circles in Figure 3) are matched by the simulation input rotation profile if we assume that the convection cells are anchored at depths equal to their widths. Shallower cells would give unstable gradients. The mass density increases nearly quadratically with depth so it is reasonable to expect the cells to be advected by the flows near their deepest extent.

This association between cell wavelength and depth indicates that the poleward meridional flow seen at the surface reverses at a depth of 35 Mm - the base of the surface shear layer where the rotation rate reaches its maximum. Although this shallow return flow violates the assumptions of flux transport dynamos \citep{DikpatiChoudhuri94, DikpatiCharbonneau99, NandyChoudhuri02, Dikpati_etal06, Choudhuri_etal07}, it was predicted by numerical simulations of the effects of rotation on supergranules \citep{Hathaway82}, it is in agreement with global helioseismology, and it helps to reconcile other observations. 

The surface has the slowest rotation and the fastest meridional flow. Small magnetic elements rotate faster than the surface and have poleward meridional flow which is slower than the surface \citep{Komm_etal93, HathawayRightmire10, HathawayRightmire11}. Both velocity components for the small magnetic elements are matched at a depth of about 15 Mm. While supergranules do have a broad range of sizes, their spectrum exhibits an excess of power at wavelengths of 30-35 Mm \citep{Hathaway_etal00}. Our results indicate that cells this size have depths roughly equal to the depth of the surface shear layer. This hardly seems coincidental but rather suggests an intimate connection between the characteristic size of supergranules and the depth of the surface shear layer.

A meridional circulation confined to the surface shear layer would also explain why numerical simulations of the solar convection zone below this surface shear layer have had difficulty producing the observed flows \citep{Miesch_etal00}. In particular, the meridional circulations in these simulations are highly structured in latitude and highly variable in time. The source of this structure and variability can be attributed to the small number ($\sim 100$) of convection cells that populate the simulated volume. A meridional circulation driven by ($\sim 10000$) supergranules - the convection cells that populate the surface shear layer - is far more likely to be less structured and variable.

This detection of a shallow equatorward return flow for the Sun's meridional circulation indicates the need for a reassessment of solar dynamo theory. The flux transport dynamo models, all of which assume and require a deep meridional flow, apparently cannot be correct. While other dynamo models exist, the majority of these place the dynamo action at the base of the convection zone with a possible secondary and less organized dynamo in the surface shear layer. Upon reassessment we may find that the surface shear layer plays a far more important role in the global dynamo.

\acknowledgements
The author would like to thank NASA for its support of this research through a grant
from the Heliophysics Causes and Consequences of the Minimum of Solar Cycle 23/24 Program to NASA Marshall Space Flight Center. He is also indebted to Ron Moore and Lisa Rightmire who read and commented on the manuscript. SOHO, is a project of international cooperation between ESA and NASA.

\end{document}